\documentclass[preprintnumbers,preprint,aps,amsmath,amssymb,amstex]{revtex4-1}
\pdfoutput=1
\usepackage{color}
\usepackage{graphics}
\usepackage{amsmath, amssymb, epsfig}
\newcommand{\mathsym}[1]{{}}

\def\10{$SO(10)$}

\newcommand{\ba}{\begin{array}}
\newcommand{\ea}{\end{array}}
\newcommand{\be}{\begin{equation}}
\newcommand{\ee}{\end{equation}}
\newcommand{\beqa}{\begin{eqnarray}}
\newcommand{\eeqa}{\end{eqnarray}}
\def\321{$SU(3)\times SU(2)\times U(1)$}
\def\b126{$\overline{126}$}
\def\mt{$\mu$-$\tau$ }
\def\mnuf{${\cal M}_{\nu f }~$}
\newcommand{\dms}  {\Delta m^2_{sol}}
\newcommand{\Dma}  {\Delta m^2_{atm}}

\begin{document}
\vspace*{1cm}
\title{Viability of the exact tri-bimaximal mixing at $M_{GUT}$ in \10}
\bigskip
\author{Anjan S. Joshipura\footnote{anjan@prl.res.in}
and Ketan M. Patel\footnote{kmpatel@prl.res.in}}
\affiliation{Physical Research Laboratory, Navarangpura, Ahmedabad-380 009,
India. \vskip 1.0truecm}

\begin{abstract}
General structures of the charged lepton and the neutrino mixing matrices leading to 
tri-bimaximal leptonic mixing are determined. These are then integrated into an $SO(10)$ model
within which detailed fits to fermion masses and mixing angles are given. It is shown that one can
obtain excellent fits to all the fermion masses and quark mixing angles keeping tri-bimaximal
leptonic mixing intact. Different perturbations to the basic structure are considered and those
which can or which cannot account for the recent T2K and MINOS results on the reactor mixing angle
$\theta_{13}^l$ are identified.

\end{abstract}


\maketitle

\section{Introduction}
The Tri-bimaximal (TBM) leptonic mixing \cite{hs}  provides an important clue in search of 
possible flavour structure \cite{af} governing the leptonic masses and mixing angles. It predicts
$\sin^2\theta_{12}^l=1/3$ and $\sin^2\theta_{23}^l=1/2$ respectively for the solar and the
atmospheric mixing angles both of which agree nearly within 1$\sigma$ with the latest global
analysis \cite{fogli, valle} of the neutrino oscillation data. TBM also predicts vanishing reactor
mixing angle $\theta_{13}^l$. This can be reconciled with the latest T2K \cite{t2k} (MINOS
\cite{minos}) results at 2.5$\sigma$ (1.6$\sigma)$ and with the global analysis \cite{fogli,valle}
at about 3$\sigma$. This suggests that the TBM may be a good zeroth order approximation which needs
perturbations affecting mainly the reactor mixing angle $\theta_{13}^l$. While such perturbations
may arise from some underlying flavor symmetry, see \cite{modelflavor} for examples, it would be
more appropriate if some independent mechanism like
grand unified theory (GUT) \cite{moh:t2k} governs these perturbations. We wish to analyze
here the TBM structure and perturbations to it within a grand unified model based on $SO(10)$ gauge
symmetry.

Incorporating  the tri-bimaximal mixing into GUTs, particularly based on the $SO(10)$ gauge group is
quite challenging, see \cite{sm} for some examples. Since all fermions in a given generation are
unified into a single $16$ dimensional irreducible representation of \10, imposition of the TBM
structure on the leptonic mass matrices  also constrains the quark mass matrices. It is not clear if
the
requirement of the exact tri-bimaximal mixing among leptons would be consistent with a precise
description of the quark masses and mixing. We suggest a general method of
incorporating the exact TBM structure within $SO(10)$ and use it to obtain quantitative description
of the fermion masses and mixing.

Our approach is purely phenomenological. We do not use any flavour symmetry but determine the most
general structure of the leptonic mixing matrices required for obtaining the tri-bimaximal mixing.
We then try to integrate this structure into \10 and discuss numerical fits to the quark masses and
mixing leaving the TBM intact. Then we discuss perturbations to the basic structure and their
effects on observables.

\section{Leptonic mixing matrices and TBM}
We shall derive general forms for the neutrino mass matrix $M_\nu$ and the left handed charged
lepton mixing matrix $U_l$ which lead in the flavour basis to a neutrino mass matrix \mnuf
exhibiting the TBM structure. We define the TBM structure for \mnuf  as follows:
\be \label{explicitmnuf}
{\cal M}_{\nu f}=
\frac{1}{3}\left(
\ba{ccc}
2 f_1+f_2&f_2-f_1&f_2-f_1\\
f_2-f_1&\frac{1}{2}(f_1+2f_2+3f_3)& \frac{1}{2}(f_1+2f_2-3f_3)\\
f_2-f_1&\frac{1}{2}(f_1+2f_2-3f_3)&\frac{1}{2}(f_1+2f_2+3f_3)\\
\ea
\right)~,
\ee
where  $f_{1,2,3}$ are complex neutrino masses. This matrix is diagonalized by
\be\label{pmns}
U_{PMNS}=O_{TBM}Q~,\ee
where $Q$ is a diagonal phase matrix and
\be \label{utbm}
O_{TBM}=\left( \ba{ccc} \sqrt{\frac{2}{3}}&\frac{1}{\sqrt{3}}&0\\
		  -\frac{1}{\sqrt{6}}&\frac{1}{\sqrt{3}}&-\frac{1}{\sqrt{2}}\\
                  -\frac{1}{\sqrt{6}}&\frac{1}{\sqrt{3}}&\frac{1}{\sqrt{2}} \\ \ea \right) \ee

A more general definition of  TBM structure would be to replace ${\cal M}_{\nu f}$ and $U_{PMNS}$
above by $P_l{\cal M}_{\nu f}P_l$ and $P_l^* U_{PMNS}$, where $P_l$ denotes a diagonal phase matrix.
Since $P_l$ can be rotated away by redefining the charged lepton fields, we shall refer to TBM
structure as the one defined by Eqs.~(\ref{explicitmnuf},\ref{pmns}).

It is known \cite{lam} that \mnuf  in Eq.~(\ref{explicitmnuf}) is invariant under a $Z_2\times Z_2$
symmetry. The elements of the $Z_2\times Z_2$ are defined as
\be \label{s2s3} \ba{cc}
S_2=\dfrac{1}{3}\left( \ba{ccc} -1&2&2\\
		  2&-1&2\\
                  2&2&-1 \\ \ea \right) ~~~{\rm and}~~~
&
S_3=\left( \ba{ccc} 1&0&0\\
		  0&0&1\\
                  0&1&0 \\ \ea \right).\ea \ee
and satisfy
\be \label{symmetrynuf}
S_{2,3}^{T}{\cal M}_{\nu f}S_{2,3}={\cal M}_{\nu f}~ \ee

We shall exploit this  $Z_2\times Z_2$ invariance in arriving at the structure of $U_l$. As noted in
\cite{ab}, one can always choose a specific basis in which $M_\nu$ exhibits the TBM structure and is
thus invariant under $Z_2\times Z_2$:
\be \label{symmetrynu}
S_{2,3}^{T}M_\nu S_{2,3}= M_\nu~  \ee
If $M_\nu'$ in an arbitrary basis is not invariant under $Z_2\times Z_2$ then one can go to a new basis
with $M_\nu=U^TM_\nu'U$ and choose $U$ in such a way that
\be \label{unumnu}
U_\nu^T M_\nu U_\nu=D_\nu~,\ee
with $D_\nu$  a diagonal matrix with real positive elements and
\be
\label{unu}
U_\nu\equiv O_{TBM} P~.\ee
$P$ being a general diagonal phase matrix. Let $U_l$ denote the mixing matrix among the left handed 
charged leptons in a basis in which $M_\nu$ is $Z_2\times Z_2$ symmetric. If such a defined $U_l$
itself is $Z_2\times Z_2$ symmetric, {\it i.e.} satisfies
\be \label{symmetryul}
S_{2,3}^{T}U_lS_{2,3}=U_l~ \ee
then \mnuf will also satisfy Eq.~(\ref{symmetrynuf}) and thus would  exhibit the TBM structure of
Eq.~(\ref{explicitmnuf}). This follows trivially from the definition
\be \label{mnuf}
{\cal M}_{\nu f}=U_l^TM_\nu U_l~, \ee
after using Eqs.~(\ref{symmetrynu},\ref{symmetryul}). Thus the $Z_2\times Z_2$ invariance of $U_l$
is sufficient to ensure the TBM for \mnuf.

Above equation allows us to determine TBM preserving class of $U_l$ in a basis with $M_\nu$
satisfying Eq.~(\ref{symmetrynu}). $S_3$ invariance corresponds to imposing the $\mu$-$\tau$
interchange symmetry on $U_l$. The $S_2$ invariance further requires $U_l^T=U_l$ and that the sum
of elements in each of its raw must be equal. Such a $U_l$ can be parameterized as
\be \label{ulp}
U_l=e^{i \alpha} P_l\tilde{U_l} P_l~,\ee
where $P_l={\rm diag.}(1,e^{i \beta},e^{i \beta})$ is a diagonal phase matrix and
\be \label{ultilde}
\tilde{U_l}=\left(
\begin{array}{ccc}
 c_\theta & \frac{s_\theta}{\sqrt{2}}  & \frac{s_\theta}{\sqrt{2}}  \\
 \frac{s_\theta}{\sqrt{2}}& -\frac{1}{2} (c_\theta + e^{i \delta})
& -\frac{1}{2} (c_\theta - e^{i \delta}) \\
 \frac{s_\theta}{\sqrt{2}}& -\frac{1}{2} (c_\theta - e^{i \delta}) & -\frac{1}{2} (c_\theta + e^{i
\delta}) \end{array}
\right), \ee
with $\tan\theta=-2\sqrt{2}\cos \beta$. $U_l$ is thus fully determined by three phase angles
$\alpha, \beta$ and $\delta$.

The form of $U_l$ as given above is the most general one required  in order to obtain  TBM  in the
basis with $M_\nu$ satisfying Eq.~(\ref{symmetrynu}). The generality is proved by noticing that the
$Z_2\times Z_2$ invariance of $U_l$ is also necessary if \mnuf is to exhibit the TBM structure.
This follows in a straightforward manner. Assume that \mnuf has TBM structure of
Eq.~(\ref{explicitmnuf}). The $U_{PMNS}$ matrix in this case can be chosen to have the form in
Eq.~(\ref{pmns}). Since $U_l=U_\nu U_{PMNS}^\dagger$, it has the following form in the basis
specified by Eq.~(\ref{unu}):
\beqa \label{ulparam}
U_l&=&O_{TBM}PQ^*O^T_{TBM}~,\nonumber \\
&=&
\frac{1}{3}\left(
\ba{ccc}
2 p_1+p_2&p_2-p_1&p_2-p_1\\
p_2-p_1&\frac{1}{2}(p_1+2p_2+3p_3)& \frac{1}{2}(p_1+2p_2-3p_3)\\
p_2-p_1&\frac{1}{2}(p_1+2p_2-3p_3)&\frac{1}{2}(p_1+2p_2+3p_3)
\ea \right)~, \eeqa
where $p_i$ denote the elements of the diagonal phase matrix $P Q^*$. Interestingly, the above $U_l$
is obtained from the general TBM \mnuf Eq.~(\ref{explicitmnuf}), by replacing the neutrino masses
with the phases $p_i$ and like \mnuf such a $U_l$ is automatically $Z_2\times Z_2$ symmetric. Thus
Eq.~(\ref{symmetryul}) also becomes necessary for the  $Z_2\times Z_2$ invariance of \mnuf.
Eq.~(\ref{ulparam}) provides an alternative parametrization of $U_l$. It reduces to the earlier
parametrization in Eq.~(\ref{ultilde}) with the definition
\beqa p_1&=&-e^{i(\alpha+\beta-\eta)},\nonumber\\
p_2&=&~e^{i(\alpha+\beta+\eta)},\nonumber\\
p_3&=&-e^{i(\alpha+2\beta+\delta)},\eeqa
with $\cos\eta=-3 c_\theta \cos\beta$ and $\sin\eta=-c_\theta\sin \beta$.

We note that the $U_\nu$ and $U_l$  in Eqs.~(\ref{unu},\ref{ulp}) are defined up to a simultaneous
redefinition $U_\nu\rightarrow U U_\nu$ and $U_l\rightarrow UU_l$. In addition, $U_l$ can also be
multiplied by an arbitrary phase matrix from the right.  Since $P_l$ is arbitrary, the $U_l$ in
some model may not appear to have the $Z_2\times Z_2$ invariance even in  a basis with  $U_\nu$
chosen as in Eq.~(\ref{unu}).  But the above exercise shows that $U_l$ can always be chosen
to have the TBM form by appropriate rephasing of the charged lepton fields.

The above reasoning can be applied to more general patterns of mixing and not just to TBM.
The key role in this construction is played by the fact that one can always choose a basis in which
$M_\nu$ is $Z_2\times Z_2$ symmetric. This follows from the fact that the $Z_2\times Z_2$ symmetry
does not put any restrictions on the neutrino masses but only on the structure of mixing. As long
as the neutrino mass matrices obey such ``mass-independent'' symmetries, the above construction of
determining the most general $U_l$ can be carried through. One can indeed define \cite{lam} an
appropriate $Z_2\times Z_2$ symmetry  corresponding to every mixing pattern  and then impose this
symmetry on $U_l$ to obtain the desired mixing structure in the flavour basis. As an example, in
case of the \mt symmetry, one can always choose a basis in which $M_\nu$ is \mt symmetric.
Then requiring that $U_l$ also be \mt symmetric, we arrive at a general forms for $U_l$ and $M_\nu$
which lead to a \mt symmetric \mnuf:\\
\be \label{mt}
\ba{cc}
M_\nu=\left(
\ba{ccc}
X&A&A\\
A&B&C\\
A&C&B\\
\ea \right)
&~;~~~
U_l=e^{i \alpha} P_{lL}\tilde{U_l} P_{lR}~
\ea
\ee
where $P_{lL,R}={\rm diag.}(1,e^{i \beta_{L,R}},e^{i \beta_{L,R}})$ and $\tilde{U_l}$ has the same
form of Eq.~(\ref{ultilde}) but now  $\theta$ is an independent parameter and is not a function of
$\beta$ as before.

\section{\10 model and TBM}

We now integrate the above leptonic structures into an \10 model. We make the following assumptions
which lead to simplification and allows us to obtain quantitative description. We (1) consider a
supersymmetric \10 model with the Higgs transforming as $10,\overline{126},120$ representations of
$SO(10)$ (2) impose the generalized parity as in \cite{moh120,grimus1,mutau,ab,jp} leading to
Hermitian mass matrices and (3) assume that the dominant contribution to $M_\nu$ is a type-II
seesaw, {\it i.e.} linear in the $\overline{126}$ Yukawa coupling. The last assumption with its
attractive consequences is made in a number of \10 models
\cite{ab,moh120,minimal,nonminimal}. {The type-II dominance may be achieved by adding more Higgs
fields.
The type-I contribution can be suppressed by pushing the $B-L$ breaking scale high while the type
-II contribution can be relatively enhanced if the Higgs triplet remains below the GUT scale. The
latter spoils the gauge coupling unification. It was pointed out  in \cite{moh54} that breaking
$SO(10)$ to $SU(5)\times U(1)$  around  $10^{17}$ GeV can suppress the type-I contribution and 
presence of $54$ plet can allow a complete multiplet transforming as 15 in $SU(5)$  to remain light.
One can achieve in this way the type-II dominance without sacrificing the gauge coupling
unification. The same phenomena was analyzed subsequently \cite{goran2} in a model with
fields transforming as $54+45$ (instead of the standard $210$) and having a $120$ plet of Higgs. The
fields
$54+45$  required to achieve the type-II dominance  do not effect the fermion masses 
and the fermion sector can be the same as considered here.}

The fermion mass relations in this case after electroweak symmetry breaking can be  written in
their most general forms as \cite{grimus1,mutau,jp}:
\beqa \label{genmass}
M_d&=& H+F+i G~,\nonumber \\
M_u&=&r (H+s F+i t_u~ G~), \nonumber\\
M_l&=& H-3 F+~it_l~ G~,\nonumber\\
M_D&=&r (H-3s F+i t_D~ G~),\nonumber\\
M_L&=& r_L F~,\nonumber\\
M_R&=& r_R^{-1} F.\eeqa
where ($G$) $H$, $F$ are real (anti)symmetric matrices. $r,s,t_l,t_u,t_D,r_L,r_R$ are
dimensionless real parameters. The effective neutrino mass matrix for three light neutrinos
resulting after the seesaw mechanism can be written as
\be \label{mnu}
M_\nu=r_LF-r_RM_DF^{-1}M_D^T\equiv  M_\nu^{II}+M_\nu^{I}
~.\ee
The first term proportional to $F$ denotes  type-II seesaw contribution. In the numerical
analysis that follows, we shall assume that $M_\nu$ is entirely given by this term and subsequently
analyze the effect of a small type-I corrections on the numerical solution found.

We can always rotate the  16-plet fermions in generation space in such a way that $M_\nu \propto F$
is diagonalized by the TBM matrix.
\be \label{basis}
F\rightarrow R^TFR=F_{TBM}\equiv O_{TBM}~{\rm Diag.}(f_1,f_2,f_3)~O_{TBM}^T
\ee
where $f_i$ are now real eigenvalues of $F$ and the $O_{TBM}$ is given by Eq.~(\ref{utbm}). The
matrix ($G$)$H$ maintains its (anti)symmetric form in such basis and we use the same label for them
in the rotated basis. In case of  type-II seesaw dominance, the light neutrino mass matrix
$M_\nu=r_L F_{TBM}$ has the form given on the RHS of Eq.~(\ref{explicitmnuf}). The model has
altogether 17 independent real parameters (3 in $F_{TBM}$, 6 in $H$, 3 in $G$, $r$, $s$, $t_u$,
$t_l$ and $r_L$) which determine the entire 22 low energy observables of the fermion mass spectrum.
Some of these parameters can be fixed by the known values of observables directly. As noted in
\cite{jp}, the \10 relation for the charged lepton mass matrix in Eq.~(\ref{genmass}) can be
rewritten as 
\be \label{mlfixing}
H+it_lG=V_l D_l V_l^\dagger+3 F_{TBM}~,\ee
where $D_l$ is a diagonal charged lepton mass matrix.
Since $H$ and $G$ are real, the real and imaginary parts of the RHS separately determine $H$ and
$t_l G$ in terms of the charged lepton masses, parameters of $F_{TBM}$ and $V_l$. $V_l$ is a
unitary matrix that diagonalizes $M_l$ and contains nine free parameters in the most general
situation. One can suitably write $V_l=\tilde{V_l}P$ where $P$ is diagonal phase matrix and
$\tilde{V_l}$ contains six real parameters. From Eq.~(\ref{mlfixing}), it is easy to see that the
phase matrix $P$ does not play any role in determining $H$ and $G$ and can be removed. So the nine
real parameters of LHS can be related to  six real parameters of $\tilde{V_l}$, three charged lepton
masses and parameters of $F_{TBM}$ in Eq.~(\ref{mlfixing}). This fixing helps us in numerical
analysis as we will see in the next subsection.

We shall present numerical analysis in two different cases. (A) Corresponding to the most general
$V_l$ (B) with $V_l=U_l$ given as in Eqs.~(\ref{ulp},\ref{ultilde}). The case (A) has already been
studied numerically in \cite{moh120,ab,jp}. We refine this analysis using a different numerical
procedure and taking into account the results of the most recent global fits \cite{fogli} to
neutrino data. This also serves as a benchmark with which to compare the case (B) which leads to
the exact TBM at $M_{GUT}$.

\subsection{Numerical Analysis: The most general case}
We study the viability of Eq.~(\ref{genmass}) with the experimentally observed values of fermion
masses and mixing angles through numerical analysis. For this, we construct a $\chi^2$ function
defined as
\be \label{chisq}
\chi^2=\sum_i \left(\frac{P_i-O_i}{\sigma_i}\right)^2 ~\ee
where the sum runs over different observables. $P_i$ denote the theoretical values of observables
determined by the expressions given in Eq.~(\ref{genmass}) and $O_i$ are the experimental values
extrapolated to the GUT scale. $\sigma_i$ denote the 1$\sigma$ errors in $O_i$. Our choice of the
input values of  quark and lepton masses and quark mixing angles are the  same as used in
\cite{ab}. In this data set, the charged fermion masses at the GUT scale are \cite{dasparida}
obtained from the low energy values using MSSM and $\tan \beta=10$. We use the input values of
lepton mixing angles from \cite{fogli} which includes results from T2K and MINOS. The effect of the
 RG evolution on the quark mixing angles is known to be negligible. This is also true for the lepton
mixing angles
in case of the hierarchical neutrino spectrum. We assume such hierarchy in neutrino masses and
therefore the input values of the quark mixing angles, CP phase and neutrino parameters we use
correspond to their values at low energy. We reproduce all these input values in Table
\ref{tab:input} for convenience of the reader.
\begin{small}
\begin{table}[ht]
\begin{math}
\begin{array}{|c|c||c|c|}
\hline
 \multicolumn{4}{|c|}{\text{GUT scale values
 with propagated uncertainty} }\\
\hline
 m_d(\text{MeV}) &1.24\pm0.41  	&\dms(\text{eV$^2$})&(7.58\pm 0.22)\times10^{-5} \\
 m_s(\text{MeV}) &21.7\pm5.2 	&\Dma(\text{eV$^2$})&(2.35\pm 0.12)\times10^{-3}\\
 m_b(\text{GeV}) & 1.06^{+0.14}_{-0.09}		& \sin  \theta _{12}^q & 0.2243\pm 0.0016 \\
 m_u(\text{MeV}) & 0.55\pm0.25		 	& \sin  \theta _{23}^q & 0.0351\pm 0.0013 \\
 m_c(\text{GeV}) & 0.210\pm0.021	& \sin  \theta _{13}^q & 0.0032\pm 0.0005 \\
 m_t(\text{GeV}) &82.4^{+30.3}_{-14.8}		& \sin ^2 \theta _{12}^l & 0.306^{+0.018}_{-0.015}
\\
 m_e(\text{MeV}) &0.3585\pm0.0003 	&\sin ^2 \theta _{23}^l&0.42^{+0.08}_{-0.03}\\
 m_{\mu }(\text{MeV})&75.672\pm0.058 	& \sin ^2\theta _{13}^l & 0.021^{+0.007}_{-0.008} \\
 m_{\tau }(\text{GeV})&1.2922\pm0.0013	&J_{CP}  &(2.2\pm0.6)\times 10^{-5}\\
\hline
\end{array}
\end{math}
\vspace{0.5cm}
\caption{Input values for quark and leptonic masses and mixing angles in the MSSM extrapolated at
$M_{GUT}=2\times 10^{16}$ GeV for $\tan\beta=10$ which we use in our numerical analysis.}
\label{tab:input}
\end{table}
\end{small}

We fit the above data to the fermion mass relations (\ref{genmass}) predicted in the model by
numerically minimizing the $\chi^2$ function. As already mentioned above, this exercise has
been done in \cite{ab} recently and a very good fit corresponding to $\chi^2_{min}=0.127$ is
found. We repeat the same analysis because of the following differences in our fitting procedure
 and because of the emergence of new results on $\theta_{13}^l$ \cite{t2k,minos}.
\begin{itemize}
 \item Compared to other observables, the charged lepton masses are known very precisely
with extremely small errors in their measurements. Instead of fitting them through $\chi^2$
minimization, we use their central values as  inputs  on the RHS of Eq.~(\ref{mlfixing}). Because of
this, our definition of the  $\chi^2$ function in Eq.~(\ref{chisq}) does not include the charged
lepton masses in it.
 \item We also use the central value of the solar to the atmospheric mass squared difference ratio
as an input and use it to fix $f_3$ through the following relation:
\be \label{f3}
f_3=f_2\left(\dfrac{\Dma}{\dms}+\left(1- \dfrac{\Dma}{\dms}\right)\left(
\dfrac{f_1}{f_2}\right)^2
\right)^{1/2}.\ee
After obtaining the solution, the overall scale of neutrino masses $r_L$ at the minimum is
determined by using the atmospheric scale as a normalization.
\end{itemize}
As a result of these simplifications, the $\chi^2$ function in our approach includes only 13
observables, 6 quark masses, 3 quark mixing angles, a CKM phase and 3 lepton mixing angles. These 
are complex nonlinear functions of 12 real parameters (2 in $F_{TBM}$, 6 in $V_l$, $r$, $s$, $t_l$
and $t_u$). This $\chi^2$ is numerically minimized using the function minimization tool MINUIT.
The results of our analysis are shown in column A in Table \ref{table:op}.
\begin{table}[ht]
\begin{small}
\begin{math}
\begin{array}{|c||c|c||c|c||c|c|}
\hline
  & \multicolumn{2}{|c||} {\text{\bf Case A}} & \multicolumn{2}{|c||} {\text{\bf Case B1}} &
\multicolumn{2}{|c|} {\text{\bf Case B2}} \\
\hline
 \text{Observables} & \text{Fitted value} & \text{Pull} & \text{Fitted value} &
\text{Pull} & \text{Fitted value} & \text{Pull} \\
\hline
 m_d [{\rm MeV}] & 1.2339 & -0.0148738 & 1.22098 & -0.0463899 & 1.02686 & -0.519852 \\
 m_s [{\rm MeV}] & 21.7214 & 0.00411949 & 21.9922 & 0.0561874 & 22.0058 & 0.058806 \\
 m_b [{\rm GeV}] & 1.06614 & 0.0438763 & 1.16345 & 0.738942 & 1.2842 & 1.60145 \\
 m_u [{\rm MeV}] & 0.550018 & 0.000073755 & 0.550234 & 0.000936368 & 0.550787 & 0.00314771
\\
 m_c [{\rm GeV}] & 0.209977 & -0.00111886 & 0.209952 & -0.00230315 & 0.210481 & 0.0229054 \\
 m_t [{\rm GeV}] & 82.5278 & 0.00421748 & 82.5855 & 0.00612198 & 81.7487 & -0.0440052 \\
 m_e [{\rm MeV}] & 0.3585 & - & 0.3585 & - & 0.3585 & - \\
 m_{\mu } [{\rm MeV}] & 75.672 & - & 75.672 & - & 75.672 & - \\
 m_{\tau }[{\rm GeV}] & 1.2922 & - & 1.2922 & - & 1.2922 & - \\
 \left( \dfrac{\dms}{\Dma}\right) & 0.0323 & - & 0.031875 & - & 0.031875 & - \\
 \sin  \theta _{12}^{q} & 0.224299 & -0.000688878 & 0.2243 & 0.0002182 & 0.224303 &
0.0019076 \\
 \sin  \theta _{23}^{q} & 0.0351032 & 0.00246952 & 0.0350951 & -0.0038047 & 0.0351294 &
0.022597 \\
 \sin  \theta _{13}^{q} & 0.00320513 & 0.0102511 & 0.00319436 & -0.0112796 & 0.0031749 &
-0.0502087 \\
 \sin ^2 \theta _{12}^{l} & 0.306119& 0.00660722 & \textbf{0.3333} & - & \textbf{0.3333} & - \\
 \sin ^2 \theta _{23}^{l} & 0.418475 & -0.0508353 & \textbf{0.5} & - & \textbf{0.5} & - \\
 \sin ^2 \theta _{13}^{l} & 0.0207708 & -0.0286467 & \textbf{0} & - & \textbf{0} & - \\
 J_{CP} & 2.19 \times 10^{-5} & -0.0183401 & 2.21\times 10^{-5} & 0.0194165 & 2.25\times 10^{-5} &
0.0845729 \\
 \delta _{MNS} & \textbf{282.396} & - & - & - & - & -\\
 \alpha _1 & \textbf{142.975} & - & \textbf{160.829} & - & \textbf{180} & - \\
 \alpha _2 & \textbf{22.4851} & - & \textbf{318.593} & - & \textbf{0} & - \\
 r_L & 6.89\times10^{-10} & -& 9.82\times10^{-10} & - &
3.53\times10^{-9} & - \\
\hline
\hline
 \chi^2_{min} &   & \textbf{0.0061} &    & \textbf{0.5519} &    & \textbf{2.8510}\\
\hline
 d_{FT} &   & 28751 &    & 88546 &    & 31171\\
d_{Data} &   & 220 &    & 200 &    & 200\\
\hline
\end{array}
\end{math}
\end{small}
\vspace{0.0cm}
\caption{Best fit solutions for fermion masses and mixing obtained assuming  type-II
seesaw dominance in the SUSY \10 model with $10+\overline{126}+120$ Higgs. Various
observables and their pulls at the minimum are shown for three different cases correspond to (A)
the general (non TBM) leptonic mixing, (B1) Exact TBM leptonic mixing with $U_l$ of
Eqs.~(\ref{ulp},\ref{ultilde}) and (B2) Exact TBM leptonic mixing with diagonal $M_l$ and $t_l=0$
(See the discussions in the text for more details). The predictions of different approaches are
shown in boldface.}
\label{table:op}
\end{table}

We obtain an excellent fit corresponding to $\chi^2_{min}=0.0061$ which is significantly better
than $\chi^2_{min}=0.127$ obtained in \cite{ab} using different procedure and different data set.
Parameters obtained for the best fit solution are shown in Appendix. All the observables are fitted
within the 0.05$\sigma$ deviation from their central values. The solution at its minimum almost 
reproduces the central value of $\sin^2\theta_{13}^l$ obtained in the global fit \cite{fogli}. 
We also determine a parameter $d_{FT}$ introduced in \cite{ab} which quantitatively measures the
amount of fine-tuning needed in the parameters for obtained fit when compared with $d_{Data}$ which
is a similar parameter obtained from the data only. From our fit, we obtain $d_{FT}\sim 2.9 \times
10^{4}$ compared to $d_{Data}\sim 220$. These parameters depend on the definition of $\chi^2$ and
since we do not include the charged lepton masses -which have very small errors- in our $\chi^2$,
both the parameters $d_{FT}$ and $d_{Data}$ are an order of magnitude smaller than \cite{ab}.
However the ratio $d_{Data}/d_{FT}\sim7.6\times10^{-3}$ obtained from our fit is almost similar to
$d_{Data}/d_{FT}\sim8\times10^{-3}$ obtained in \cite{ab} which shows that both of the solutions
need substantial level of fine tunings in the model parameters.

\subsection{Numerical Analysis: Exact TBM}
After discussion of the above general case, we now specialize to the case of the exact TBM. This
case is of considerable theoretical interest since it can point to some underlying symmetry existing
at $M_{GUT}$. We can implement the exact TBM in a model independent way by choosing $V_l=U_l$ in
Eq.~(\ref{mlfixing}). With this choice, all the leptonic mixing angles get fixed to their TBM
values. Also the central value of the ratio of the solar to the atmospheric (mass)$^2$ differences
is used as input and a parameter $r_L$ is determined at the minimum by using the atmospheric scale.
Thus the $\chi^2$ function in Eq.~(\ref{chisq}) now involves only observables in the quark sector.
As already discussed, $U_l$ in Eq.~(\ref{ulp},\ref{ultilde}) is parameterized by three phase angles
$\alpha, \beta$ and $\delta$. An overall phase $\alpha$ is irrelevant for the physical observables
and can be removed. This leaves only 8 real parameters (2 in $F_{TBM}$, 2 in $U_l$, $r$, $s$, $t_l$
and $t_u$) which are fitted to the 10 observables in the quark sector by minimizing the $\chi^2$.
The results are shown in column B1 in Table \ref{table:op}. The obtained fit corresponds to
$\chi^2_{min}=0.552$ ($\chi^2_{min}/{\rm d.o.f.}=0.276$). Only the fitted value of $m_b$ deviates
slightly from the central value with a $0.74\sigma$ pull. All the remaining observables are fitted
within $0.06\sigma$. A set of parameters obtained for this solutions are shown in Appendix. The fit
obtained here is not significantly different from the general case discussed before showing that all
the fermion masses and mixing angles can be nicely reproduced along with the exact TBM within the
\10 framework discussed here.

Before we discuss possible perturbations in TBM  pattern, let us discuss a very special
case corresponding to a diagonal $M_l$. This corresponds to  $U_l$ coinciding with  an identity matrix 
and is a special case of Eq.~(\ref{ulp},\ref{ultilde}) with $\beta=\frac{\pi}{2}$ and $\delta=0$.
Since $M_l$ is real and diagonal in this case, $t_l G$ must vanish in Eq.~(\ref{mlfixing}). If
$G=0$ then the quark mass matrices also become real and there is no room for CP violation. The
viable scenario must therefore have nonzero $G$ and hence $t_l=0$. As a result, unlike before, the
three parameters in $G$ do not get determined from $M_l$, see, Eq.~(\ref{mlfixing}) and remain
free. They can be fitted from the quark sector observables. We carried out a separate
numerical analysis for this particular case and the results are shown in column B2 in Table
\ref{table:op}. The fit obtained gives relatively large $\chi^2_{min}=2.85$ ($\chi^2_{min}/{\rm
d.o.f.}=1.43$) with more than 1$\sigma$ deviation in the bottom quark mass. Although the obtained
$\chi^2_{min}$ is statistically acceptable at 90\% confidence level, it is not as good as the
previous one and we shall not consider this case with $U_l=I$ any further.

\section{Perturbed TBM}
The TBM is an ideal situation and various perturbations to this can arise in the model. We need to
analyze these perturbations in order to distinguish this case from the generic case without the
built in TBM. A deviation from tri-bimaximality can arise due to
\begin{enumerate}
 \item renormalization group evolution (RGE) from $M_{GUT}$ to $M_Z$.
 \item small contribution from the  sub dominant type-I seesaw term in Eq.~(\ref{mnu}).
 \item the breaking of the $Z_2\times Z_2$ symmetry in $U_l$ which ensured TBM.
\end{enumerate}
The effect of (1) is known to be negligible \cite{rg}  in  case of the  hierarchical neutrino mass
spectrum which we obtain here. We quantitatively discuss the implications of the other two
scenarios via detailed numerical analysis in the following subsections.

\subsection{Perturbation from type-I seesaw}
Depending on the GUT symmetry breaking pattern and parameters in the superpotential of the theory, a
type-I seesaw contribution can be dominant or sub dominant compared to type-II but it is always
present and can generate deviations in an exact TBM mixing pattern in general. In the approach
pursued here it is assumed that such contribution remains sub dominant and generates a small
perturbation in dominant type-II spectrum. Eq.~(\ref{mnu}) can be rewritten as
\be \label{mnu1}
M_\nu=r_L (F - \xi M_DF^{-1}M_D^T)
~\ee
where $\xi = r_R/r_L$ determine the relative contribution of type-I term in the neutrino mass
matrix.

\begin{figure}[ht]
 \centering
 \includegraphics[width=16cm]{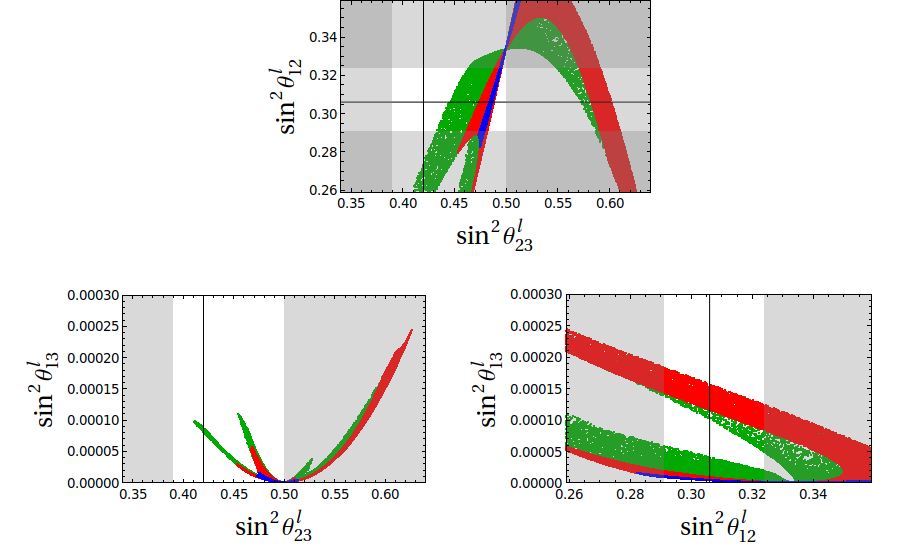}
\caption{Correlations among the lepton mixing angles when
two real parameters $\xi$ and $t_D$ are varied randomly. The points with different colors correspond to $|t_D| < 1$
(blue), $1 \le |t_D| < 5$ (red) and $5 \le |t_D| < 10$ (green). The black lines are the
updated central values of $\sin^2\theta_{12}^l$ and $\sin^2\theta_{23}^l$ obtained by the global
fits on neutrino oscillation data \cite{fogli}. The unshaded and the shaded regions
correspond to $1\sigma$ and $3\sigma$ bounds respectively.}
\label{fig:1}
\end{figure}

The second term in Eq.~(\ref{mnu}) brings in two new parameters $\xi$ and $t_D$ present in the
definition of $M_D$ in Eq.~(\ref{genmass}). These parameters however affect  only the neutrino
sector. We isolate the effect of type-I contribution by choosing other parameters at the
$\chi^2$ minimum found in Section III-B. $\xi$ and $t_D$ remain unconstrained at this minimum
and their values do not change the $\chi^2$ obtained earlier since the latter contains only the
observables in the quark sector. The $\xi,t_D$ however generate departure from the exact TBM.  We
randomly vary the parameters $\xi$ and $t_D$ and evaluate the neutrino masses and mixing angles.
While doing this, we take care that all  these observables remain within their present 3$\sigma$
\cite{fogli} limits. Such constrains allow very small values of $|\xi| \le 10^{-7}$. The
correlations between different leptonic mixing angles found from such analysis are shown
in Fig.~(\ref{fig:1}).

It is seen from Fig.~(\ref{fig:1}) that the perturbation induced by type-I term cannot generate
considerable deviation in the reactor angle if the other two mixing angles are to remain within
their 3$\sigma$ range. In particular, requiring that $\sin^2\theta_{12}^l$ remains within the
3$\sigma$ range puts an upper bound $\sin^2\theta_{13}^l \le 0.0002$ which does not agree with the
latest results from T2K and MINOS showing that a small perturbation from type-I term cannot be
consistent with data when the type-II term displays exact TBM.

\subsection{Perturbation from the charged lepton mixing}

\begin{figure}[ht]
 \centering
 \includegraphics[width=16cm]{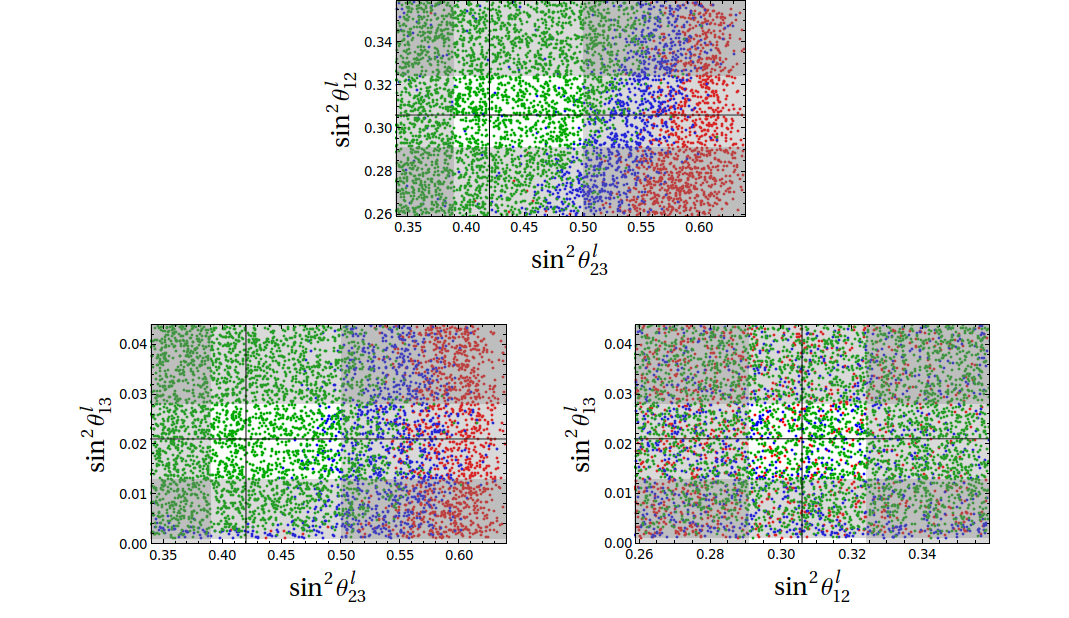}
\caption{Correlations among the lepton mixing angles in  case of the most general
charged lepton mixing matrix $U_l$. The points with different colors correspond to
$\bar{\chi}^2_{min}
< 1$ (green), $1 \le \bar{\chi}^2_{min} < 4$ (blue) and $\bar{\chi}^2_{min} \ge 4$ (red).
The black lines are the updated central values of $\sin^2\theta_{12}^l$ and $\sin^2\theta_{23}^l$
obtained by the global fits on neutrino oscillation data \cite{fogli}. The unshaded and the
shaded regions correspond to $1\sigma$ and $3\sigma$ bounds respectively.}
\label{fig:2}
\end{figure}

A different class of perturbation to TBM arise when $U_l$ deviates from its $Z_2\times Z_2$
symmetric form given in Eq.~(\ref{ulp}). In this case, the neutrino mass matrix has TBM structure
but the charged lepton mixing leads to departure from it. This case has been considered in the
general context \cite{leptonic} as well as in $SO(10)$ context \cite{ab}. Within our approach,
we can systematically look at the perturbations which change the values of any one or more angles
from the TBM value. For example, we can choose $U_l$ as given in Eq.~(\ref{mt}) and look at the
quality of fits in this case compared to the exact TBM solution. This choice (corresponding to \mt
symmetry) leaves $\theta_{23}^l$ and $\theta_{13}^l$ unchanged but perturbs $\theta_{12}^l$.
Alternative possibility is to simultaneously perturb all three mixing angles and look at the quality
of fit compared to the exact TBM case. We follow this approach. For this we choose $U_l$ to be a
general unitary matrix and go back to the analysis in Section III-A. There, we have fitted the
solar and the atmospheric mixing angles to their low energy values given in Table \ref{tab:input}.
Here, we modify the definition of $\chi^2$ and pin down a specific value $o_i$ of the mixing
angles $p_i$   by adding a term 
\be \label{chisq1}
\chi^2_{lm}=\sum_i \left(\frac{p_i-o_i}{0.01 o_i}\right)^2 ~\ee
to $\chi^2_{q}$ that contains all the observables of the  quark sector. Sum in Eq.~(\ref{chisq1})
runs over the three lepton mixing angles. The $\chi^2 = \chi^2_{q} + \chi^2_{lm}$ is then
numerically minimized to fit the 13 observables determined in terms of 12 real parameters as
mentioned in Section III-A. Artificially introduced small errors in Eq.~(\ref{chisq1}) fix the value
$o_i$ for $p_i$ at the minimum of the $\chi^2$. We then look at the quantity 
\be \label{chibar}
\bar{\chi}^2_{min}\equiv \chi^2|_{min} - \chi^2_{lm} |_{min}
\ee 
which represents the fit to the quark spectrum when the lepton mixing angles $p_i$ are pinned down
to values $o_i$. We repeat such analysis by randomly varying $o_i$ within the allowed 3$\sigma$
ranges of lepton mixing angles \cite{fogli}. The results are displayed in Fig.~(\ref{fig:2}).

We plot the correlations among the lepton mixing angles and show the corresponding values of
$\bar{\chi}^2_{min}$ in three different regions. The points corresponding to $\bar{\chi}^2_{min}
< 1$ (green) represent very good fit in which all the observables are fitted within 1$\sigma$. The
obtained fit shown by the points corresponding to $1 \le \bar{\chi}^2_{min} < 4$ (blue) is not as
good as the previous one but it is statistically acceptable. The points for $\bar{\chi}^2_{min} > 4$
(red) represent poor fit and can be ruled out at 95\% confidence level. Fig.~(\ref{fig:2}) shows
definite correlations between $\theta_{23}^l$ and $\theta_{12}^l$. It is seen that the region
$\bar{\chi}^2_{min} < 4$ falls largely below $\sin^2\theta_{23}^l= 0.58$ for $\sin^2\theta_{12}^l
= 0.306$. It is also seen from the figure that the entire range $0.001\leq\sin^2\theta_{13}^l\leq
0.044$ is consistent with statistically acceptable fits to fermion spectrum. This is to
be contrasted with the previous case where perturbation from type-I seesaw term led to an upper
bound. The bounds obtained numerically allows us to clearly distinguish the case of the exact
TBM at $M_{GUT}$ in comparison to the one in which the charged leptons lead to departures from the
tri-bimaximality. 

\section{Summary}
The presently available information on the leptonic mixing may be described by a TBM structure
for \mnuf combined with appropriate perturbations generating relatively large $\theta_{13}^l$. We
have analyzed the viability of this scenario in a larger context of the grand unified $SO(10)$
theory taking a specific model as an example. The TBM structure for the neutrino mass matrix is a
matter of choice of the basis \cite{ab}. Thus the existence of TBM is linked to the structure of the
charged lepton mixing matrix $U_l$  in this basis. All the related studies in this context
\cite{leptonic,ab} assume that $U_l$ deviates slightly from identity and discuss the breaking of the
TBM pattern through such $U_l$. We have shown that it is possible to construct a class of
non-trivial $U_l$ quite different from identity which preserve the TBM structure of $M_\nu$
when transformed to the flavour basis.  Identification of such non-trivial $U_l$ becomes crucial in
the context of $SO(10)$ and allows  us  to obtain a viable fit to fermion spectrum keeping TBM
intact. The quality of fit obtained in this case is excellent as shown in Table \ref{table:op} and
differs only marginally from a general situation without imposing the TBM structure at the outset.

The existence of TBM at the GUT scale may be inferred by considering its breaking which can arise in
the model and the reactor mixing angle is a good pointer to this. The quantum corrections are known
\cite{rg} to lead to very small $\theta_{13}^l$ for the hierarchical neutrinos. Similarly,
corrections coming from type-I seesaw term imply an upper bound, $\sin^2\theta_{13}^l\le
0.0002$ as discussed in Section IV-B. These two cases can be ruled out by relatively large value
of $\theta_{13}^l$ as indicated by the observations from T2K and MINOS. These cases are in sharp
contrast to a situation in which one does not impose the TBM at $M_{GUT}$ and determined $U_l$ from
a detailed fits to fermion masses. We have analyzed this scenario using the results from the
latest \cite{fogli} global fits to neutrino oscillation data. It is found that the entire range
$0.001\leq\sin^2\theta_{13}^l\leq 0.044$ is consistent with the detailed description of all
the fermion masses and mixing angles.\\

\noindent{\bf Acknowledgements}\\
Computations needed for the results reported in this work were done using the PRL 3TFLOP cluster at
Physical Research Laboratory, Ahmedabad. ASJ thanks the Department of Science and Technology,
Government of India for support under the J. C. Bose National Fellowship programme, grant no.
SR/S2/JCB-31/2010.\\

\section{Appendix}

In this Appendix, we show the parameter values for each of the two cases (A) and (B1) in
Table \ref{table:op}. The given values of the matrices $H$, $F$, $G$ and parameters $r,~s,~t_l$
and $t_u$ are at the minimum of $\chi^2$ and obtained from our fitting procedure. All the
physical mass matrices can be constructed from the given numbers using the \10 relations in
Eq.~(\ref{genmass}). 

\subsection*{Case A: The most general $V_l$}
The parameter values corresponding to the best fit solution shown in column A in
Table \ref{table:op} are the following.
\begin{small}
\beqa \label{prmtA}
H&=&\left(
\begin{array}{ccc}
 0.00286198 & 0.00178975 & -0.0399313 \\
 0.00178975 & 0.0196685 & 0.00741417 \\
 -0.0399313 & 0.00741417 & 1.10192
\end{array}
\right){\rm GeV} \nonumber \\
F&=&\left(
\begin{array}{ccc}
 -0.00235758 & -0.0053869 & -0.0053869 \\
 -0.0053869 & -0.0394516 & 0.0317071 \\
 -0.0053869 & 0.0317071 & -0.0394516
\end{array}
\right){\rm GeV} \nonumber \\
G&=&\left(
\begin{array}{ccc}
 0 & -0.00144676 & -0.00442083 \\
 0.00144676 & 0 & -0.0183052 \\
 0.00442083 & 0.0183052 & 0
\end{array}
\right){\rm GeV} \nonumber\\
(r,~s,~t_u,~t_l) &=& (76.2076,~0.54487,~-0.870094,~-14.5065) \nonumber  \eeqa   \end{small}

\subsection*{Case B: Exact TBM lepton mixing ($V_l=U_l$)}
The parameter values corresponding to the best fit solution shown in column B1 in
Table \ref{table:op} are the following.
\begin{small}
\beqa \label{prmtB}
H&=&\left(
\begin{array}{ccc}
 0.00111546 & 0.000428794 & -0.020097 \\
 0.000428794 & 0.0188148 & 0.0639684 \\
 -0.020097 & 0.0639684 & 1.17731
\end{array}
\right){\rm GeV} \nonumber \\
F&=&\left(
\begin{array}{ccc}
 -0.0016484 & -0.00375321 & -0.00375321 \\
 -0.00375321 & -0.0276748 & 0.0222732 \\
 -0.00375321 & 0.0222732 & -0.0276748
\end{array}
\right){\rm GeV} \nonumber \\
G&=&\left(
\begin{array}{ccc}
 0 & -0.00278749 & -0.0378996 \\
 0.00278749 & 0 & -0.0820076 \\
 0.0378996 & 0.0820076 & 0
\end{array}
\right){\rm GeV} \nonumber \\
(r,~s,~t_u,~t_l) &=& (70.5742,~0.526782,~0.551405,~2.15986) \nonumber  \eeqa   \end{small}

\end{document}